# Improvement in RF Curves for Booster Running at High Intensities

Xi Yang and Rene Padilla

*Fermi National Accelerator Laboratory*

Box 500, Batavia IL 60510

## Abstract

A feed-forward ramp can be implemented in Booster to compensate the beam energy loss at different beam intensities for the purpose of minimizing the radial error signal. This can be done only when we have a good understanding about the dependence between the beam energy loss per turn and the beam intensity experimentally. Besides, based upon this understanding we can predict the required accelerating voltage at the transition crossing for different beam intensities, which can be extremely helpful for Booster running at higher beam intensities than ever before.

## Introduction

The proton beam is accelerated from 400 MeV to 8 GeV in Booster in a time of 33.3 ms, while the RF frequency sweeps from 37.8 MHz to 52.9 MHz. The low-level RF system (LLRF) together with the bias running closed loop maintains the correct phase relation between the circulating beam bunches (CB) and the accelerating gap voltage.[1] The RF phase angle is continuously adjusted to maintain the required rate of energy gain. LLRF includes two feedback systems; phase feedback (fast feedback) keeps a constant phase relation between the CB and the low-level frequency signal (LO-Freq), which is generated by the VXI LLRF,[2] and radial feedback (slow feedback) keeps the beam at the correct radial position. The bias running closed loop maintains the minimum phase error between the RF drive and the cavity field.

The LLRF uses four approximate programmed curves to control the acceleration process: FREQ, Bias, ROF, and RAG. With the input of the phase error from phase feedback of LLRF, VXI LLRF generates the LO-Freq using the FREQ program in such a way that the LO-Freq has the same frequency and a constant phase difference with the CB through a cycle. The radial feedback system takes the signal from the radial position (RPOS) and compares it with the radial offset (ROF), which specifies the desired radial position through the cycle. The phase shift drive signal (PSDRV) is the sum of BDOT, which is the differentiation of the magnet ramp, and the product of the radial gain (RAG) and the radial error signal, which is the difference between RPOS and ROF. Phase shifter shifts the phase of the LO-Freq at the amount of PSDRV before the LO-Freq is sent to the high-level RF system (HLRF) as the RF drive. The Bias programs the bias supply current in order to keep the cavity field following the RF drive.

## System Analysis

Phase detector of the LLRF phase feedback is continuously comparing the phase difference between the CB signal from the long-18 gap detector and the LO-Freq before it is delivered to the phase shifter, and the phase difference is sent to VXI LLRF as the phase error for the correction of the LO-Freq phase through a cycle. Since the bandwidth of the LLRF phase detector is 1 MHz, the LO-Freq before the phase shifter is being kept at a constant phase relation with the CB. PSDRV ($V_{psdrv}$) in unit of volt can be calculated using eq.1 with a calibration constant of 9°/V.

$$V_{psdrv}(t) = a \cdot \frac{dB(t)}{dt} + \left(R_{pos}(t) - R_{of}(t)\right) \cdot R_{gain}(t). \qquad 1$$

Here, $B(t)$ is the magnet-ramp value at time $t$ in a cycle. $dB(t)/dt$ (BDOT) is a sine-wave with a frequency of the Booster rep rate 15 Hz. The amplitude of BDOT can be set to one since its contribution to PSDRV can be adjusted experimentally via the multiplier $a$. $R_{pos}(t)$ is the radial position of the CB from the long-18 beam position monitor (BPM), $R_{of}(t)$ is the desired radial position via the ROF curve, and $R_{gain}(t)$ is the radial gain via the RAG curve.

RFSUM is the vector sum of the RF accelerating voltages from all the cavities in the ring. The synchronous phase, which is the phase of the beam centroid relative to the sine wave of RFSUM,[3] can be calculated from PSDRV in the desired situation that the phase difference between the cavity field and the RF drive is zero, as shown in eq.2.

$$\phi_s(t) = 9 \cdot V_{psdrv}(t) = 9 \cdot \left( a \cdot \frac{dB(t)}{dt} + \left( R_{pos}(t) - R_{of}(t) \right) \cdot R_{gain}(t) \right). \qquad\qquad 2$$

Here, $\phi_s(t)$ is the desired synchronous phase at time $t$ in unit of degree.

Since PSDRV can be measured for each Booster cycle, the desired synchronous phase can be calculated using eq.2. The desired synchronous phase ($\phi_s(t)$) and measured RFSUM ($V(t)$) can be used to calculate the effective accelerating voltage ($V(t) \times \mathrm{Sin}(\phi_s(t))$). Also, the effective accelerating voltage is the sum of the accelerating voltage required by BDOT and the energy loss per beam turn. The accelerating voltage required by BDOT is known from the magnet ramp.[3] So the energy loss per beam turn can be estimated from the difference between the effective accelerating voltage and the accelerating voltage required by BDOT. The energy loss per beam turn can be obtained as a function of the beam intensity through a cycle using the linear approximation after PSDRV is measured at several beam intensities. Finally, one can predict what PSDRV should be at a known beam intensity.

## Example and Prediction

RFSUM and PSDRV were measured at extracted beam intensities of $0.4 \times 10^{12}$ protons and $4.7 \times 10^{12}$ protons, and they are shown in Figs. 1(a) and 1(b). The black and red curves in each plot represent the results for extracted beam intensities of $0.4 \times 10^{12}$ protons and $4.7 \times 10^{12}$ proton respectively. The desired synchronous phase is calculated from the PSDRV data, as shown in Fig. 1(b), using eq.2, and the result is shown in Fig. 1(c). The accelerating voltage per beam turn required by BDOT is shown in Fig. 1(d).[3] The effective accelerating voltage is calculated using "$V(t) \times \mathrm{Sin}(\phi_s(t)$" from RFSUM ($V(t)$), as shown in Fig. 1(a), and the desired synchronous phase ($\phi_s(t)$), as shown in Fig. 1(c), and the result is shown in Fig. 1(e). The energy loss per beam turn is estimated from the difference between the effective accelerating voltage, as shown in Fig. 1(e), and the accelerating voltage required by BDOT, as shown in Fig. 1(d), and the result is shown in Fig. 1(f). The linear dependence between the beam energy loss per turn and the beam intensity can be estimated using eq.3.

$$k(t) = \frac{V_{eff}(t, N_2) - V_{eff}(t, N_1)}{\left((N_2 - N_1)\big/10^{12}\right)}. \qquad\qquad 3$$

Here, $V_{eff}(t, N)$ is the effective accelerating voltage at time $t$ in a cycle for the extracted beam intensities of $N$ protons, and $k(t)$ is the linear coefficient of the dependence between the beam energy loss per turn and the beam intensity. The difference of the red curve and the black curve in Fig. 1(e) is the same with the difference in Fig. 1(f). $N_2 = 4.7 \times 10^{12}$ is the extracted beam intensity of the red curve, and $N_1 = 0.4 \times 10^{12}$ is the extracted beam intensity of the black curve. The linear coefficient $k(t)$ is calculated using eq.3, and the result is shown in Fig. 1(g). Finally, the effective accelerating voltage for the extracted beam intensity $N$ can be predicted using eq.4.

$$V_{eff}(t, N) = k(t) \cdot \left(\frac{N - N_1}{10^{12}}\right) + V_{eff}(t, N_1). \qquad\qquad 4$$

In the situation of $N = 6.5 \times 10^{12}$, the effective accelerating voltage is calculated using eq.4, and the result is shown as the blue curve in Fig. 2(a). The black curve in Fig. 2(a) is the same as the black curve in Fig. 1(a).

## Comment

From the approximation of the linear dependence between the beam energy loss per turn and the beam intensity, the energy loss for different beam intensities through a cycle can be predicted, and it can be applied as a feed-forward ramp to PSDRV for the purpose of compensating the intensity-dependent beam energy loss and minimizing the radial error signal. In the situation of the extracted beam intensity of $6.5 \times 10^{12}$, as shown in Fig. 2(a), the beam needs more RF accelerating voltage at the transition crossing than it gets now. This can be achieved by adjusting the attenuation to the Anode program before it is delivered to the modulators of RF stations. The setup of the attenuation to the Anode program is for protecting the power amplifiers of RF stations.

Another approach for getting more effective accelerating voltage at the transition crossing is to find a way to decrease the energy loss of the high intensity beam. From the past experience, the injection mismatch blew the beam emittance and made the energy loss smaller than that in the matched situation, as shown in Fig. 2(b). Both the black and red curves are at the extracted beam intensity of $4.7 \times 10^{12}$. The red curve represents the injection mismatched situation with a smaller PSDRV signal than the black curve, which represents the matched injection situation. Here, the smaller the PSDRV signal is, the less the effective accelerating voltage is. However, the aperture limit to the allowed increase in the beam emittance should also be considered, especially for the high intensity situation.

### Acknowledgement

Special thanks should be given to Bill Pellico for his help in understanding the LLRF issues related to this work.

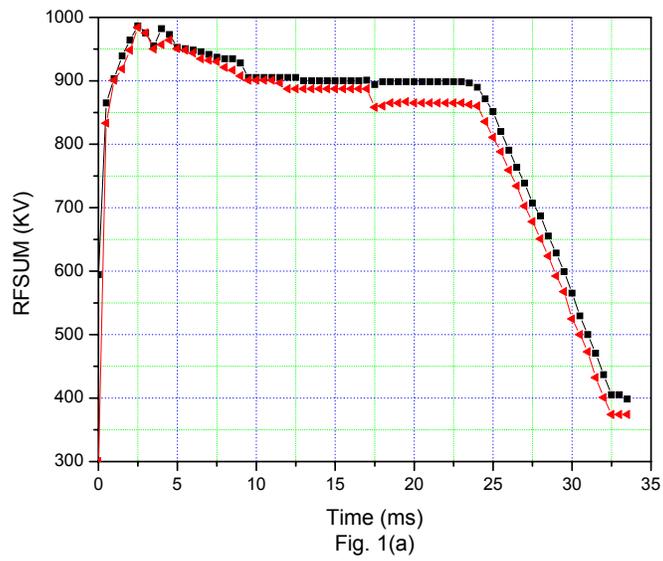

Fig. 1(a)

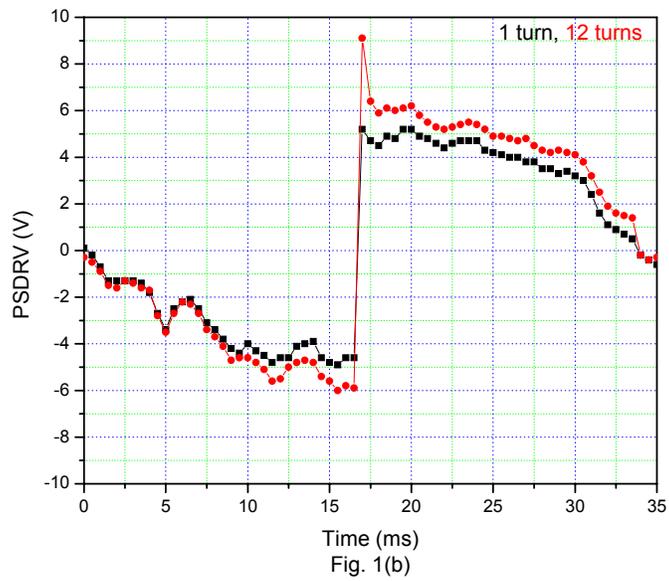

Fig. 1(b)

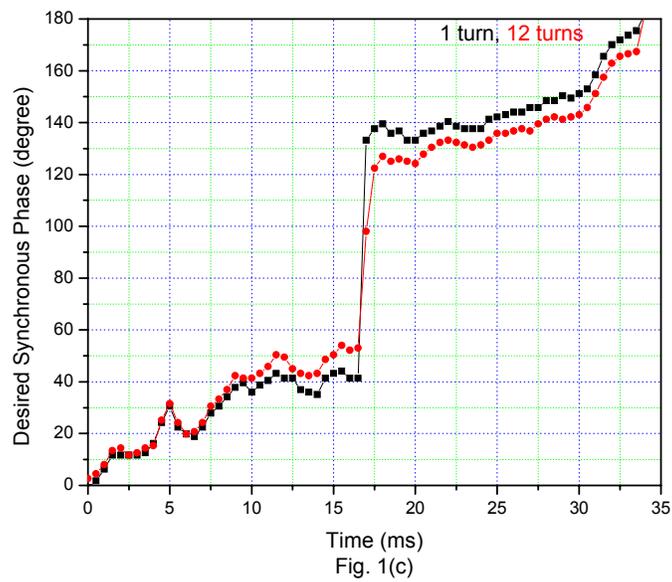

Fig. 1(c)

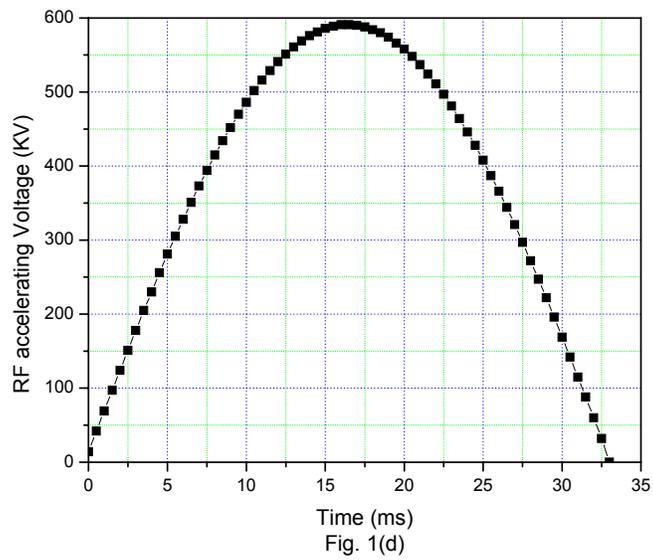

Fig. 1(d)

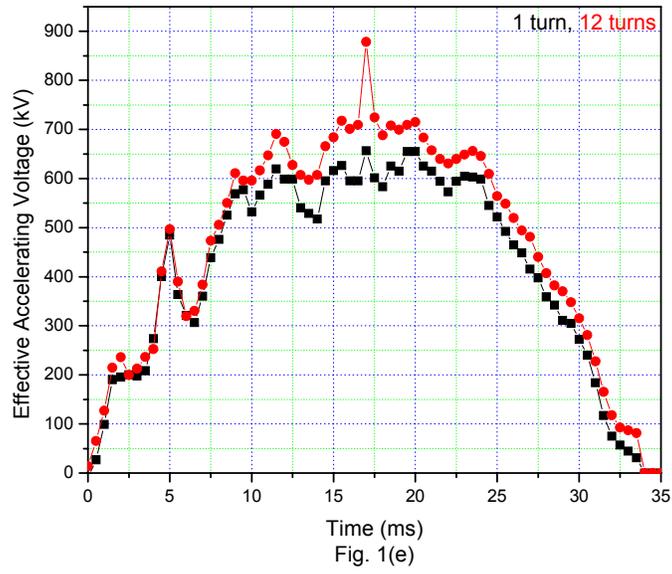

Fig. 1(e)

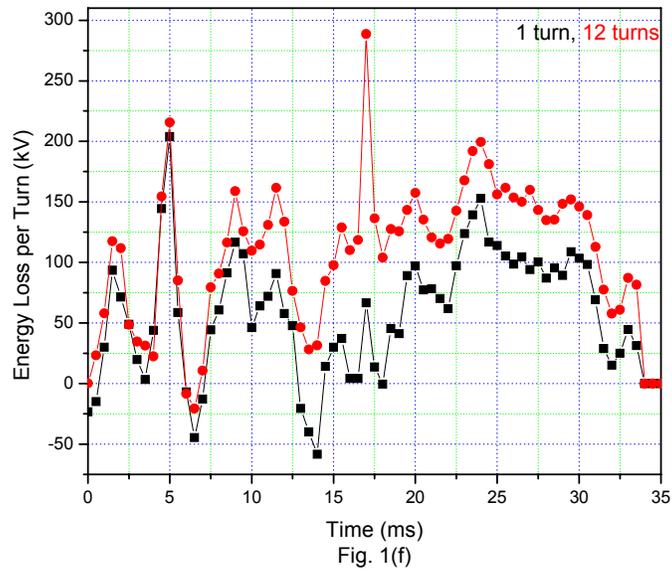

Fig. 1(f)

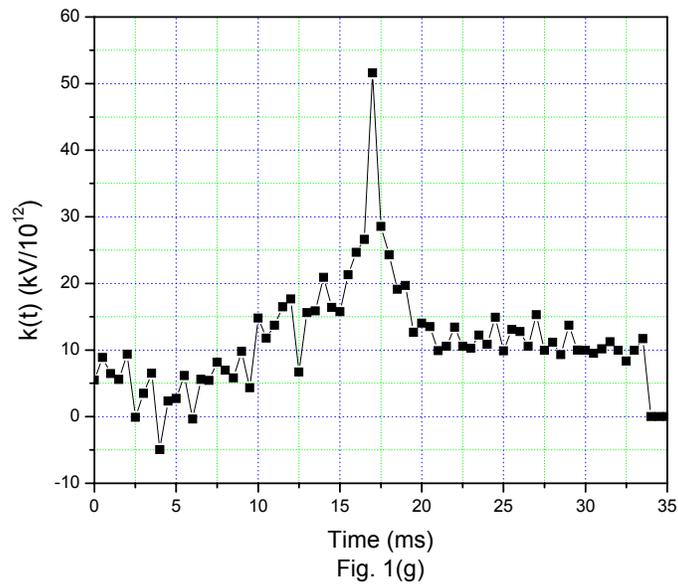

Fig. 1(g)

Fig. 1(a) RFSUM measured at extracted beam intensities of $0.4 \times 10^{12}$ protons (black curve) and $4.7 \times 10^{12}$ protons (red curve).

Fig. 1(b) PSDRV measured at extracted beam intensities of $0.4 \times 10^{12}$ protons (black curve) and $4.7 \times 10^{12}$ protons (red curve).

Fig. 1(c) desired synchronous phase for extracted beam intensities of $0.4 \times 10^{12}$ protons (black curve) and $4.7 \times 10^{12}$ protons (red curve).

Fig. 1(d) the accelerating voltage per beam turn required by BDOT in a Booster cycle.

Fig. 1(e) the effective accelerating voltage for extracted beam intensities of $0.4 \times 10^{12}$ protons (black curve) and $4.7 \times 10^{12}$ protons (red curve).

Fig. 1(f) the energy loss per beam turn estimated for extracted beam intensities of $0.4 \times 10^{12}$ protons (black curve) and $4.7 \times 10^{12}$ protons (red curve).

Fig. 1(g) the linear dependence between the beam energy loss per turn and the beam intensity estimated for a Booster cycle.

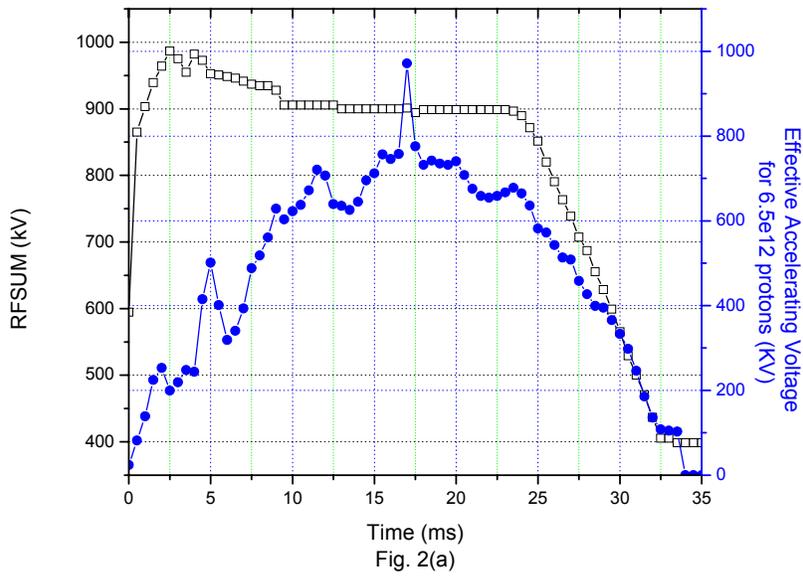

Fig. 2(a)

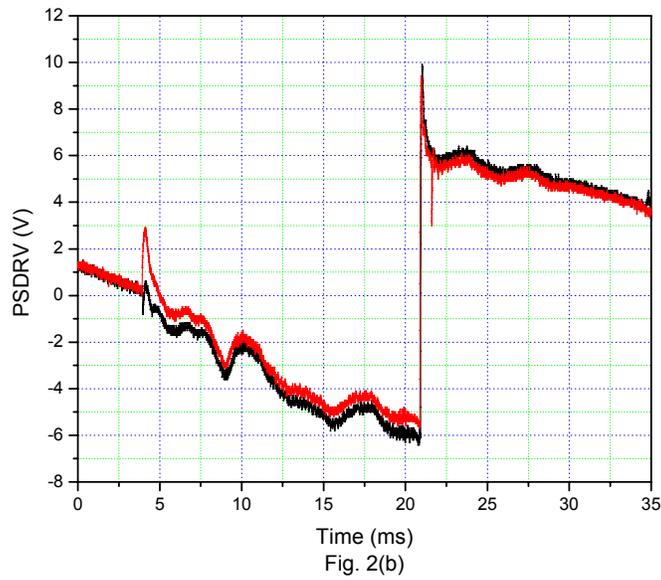

Fig. 2(b)

Fig. 2(a) the effective accelerating voltage calculated at the extracted beam intensity of $6.5 \times 10^{12}$.

Fig. 2(b) PSDRV measured at the extracted beam intensity of $4.7 \times 10^{12}$. The black curve represents the situation with a matched injection, and the red curve represents the situation with a mismatched injection.